\documentclass[12pt]{article}
\usepackage{latexsym,epsfig}
\usepackage{graphicx}
\usepackage{pifont}
\usepackage{verbatim}
\def\nn{\nonumber}
\def\ed{\end{document}}

\def\sk{\smallskip}

\def\bg{\bigskip}
\def\beq{\begin{eqnarray}}
\def\eq{\end{eqnarray}}
\def\beqn{\begin{eqnarray*}}
\def\eqn{\end{eqnarray*}}
\def\nl{\noindent}

\begin{document}
\begin{center}
{\bf \large Bounds for $Z^\prime$ Mass in 3-3-1 Models}
\bg 

{\bf \large  from $e^+$ $e^-$ Collisions at ILC and CLIC Energies}
\vskip .8 cm 

{E. Ramirez Barreto, Y. A. Coutinho}

{Universidade Federal do Rio de Janeiro}
\vskip .5 cm
 
{J. S\'a Borges}

{Universidade do Estado  do Rio de Janeiro}
\vskip .5 cm

{Rio de Janeiro - Brazil}
\end{center}

\begin {abstract}
 We obtain bounds for the mass of the extra neutral gauge boson,  ${Z^\prime}$, predicted by two versions of ${SU (3)_C \times SU (3)_L \times  U (1)_X }$ model, namely  one  corresponding to its minimal version and another where the neutrinos are allowed to have right-handed projection. We explore $\sqrt s$ from $0.5$ TeV to $5$ TeV region, that will be accessible in next linear colliders (ILC and CLIC). We used the process $e^+ + e^- \longrightarrow f^+ + f^-$, with $f=\mu$ , $c$ and $b$ to obtain the energy dependent lower bounds for $M_{Z^\prime}$, within $95\%$ C.L., by performing  a $\chi ^{2}$ fit of the difference between the final lepton angular distribution calculated from the SM and that predicted by the 3-3-1 models, for a zero mixing angle.
In addition we show that, as the angular distributions depends on $Z^\prime$ couplings,  the obtained bounds allow one for disentangle the studied models.
Finally, using the model with right-handed neutrinos,  we calculate the  total cross section for $e^+ + e^- \longrightarrow e^+ +  e^- + e^+ +  e^-$.  We concluded that  $Z^\prime$ contribution is very small when compared with the bilepton contribution of the minimal version  and with the SM cross section. 
\end{abstract}


elmer@if.ufrj.br, yara@if.ufrj.br, saborges@uerj.br
\newpage
\section{Introduction}
The Standard Model (SM) of strong and electroweak interactions is
extremely successful . Its predictions from calculations both at tree level as higher order corrections are consistent with all available experimental data. However, as the new colliders generation will explore TeV energy regime, there exists the possibility of new findings. High-energy $e^+ e^-$ linear colliders will complement and specify the probes tested in the future $p p$ collider (LHC). The common analysis of linear and hadron colliders will probably show the existence and properties of new particles and then select what are the models beyond the Standard one that can explain the new results.
On the other hand, the  theoretical extensions of the SM are motivated  by attempting to understand features that are accommodated in the SM but not totally explained by it.

The new energy scale, to be explored in future accelerators, calls for some SM extensions or alternative models.  Many proposed models predict the existence of an extra heavy neutral gauge boson,  $Z^\prime$. These include: 3-3-1 models \cite{PIV,FRA,RHN,TON}, little Higgs model \cite{LIT}, left-right symmetric models \cite{LRM}, superstring inspired E$6$ model \cite{E6M} and models with extra dimensions as Kaluza-Klein excitations of neutral gauge bosons \cite{RIZ}. 

The search for limits on $Z^\prime$ unknown properties: mass, width,  mixing angle and on its couplings with ordinary matter are both direct and indirect \cite{LEI}. Certainly the best possibility to get direct information on its parameters corresponds to the case when the energy of the collider is $\sqrt s \simeq M_{Z^\prime}$. On the other hand, the indirect determination of $Z^\prime$ properties, occurs when $\sqrt s < M_{Z^\prime}$. Since high energy observables are yet predicted from well known SM parameters, the information about  $Z^\prime$ can be extracted from  differences between SM  predictions and experimental measurements. 
       
Indirect information on new particles are related to deviations from SM data on observables, by their contributions to electroweak vacuum polarization diagrams. Vacuum polarization affects weak observables by modifying the gauge boson propagators, and are called {\it oblique corrections}. At present, oblique radiative parameters $S$, $T$, and $U$ can be used to indicate the effects of the new physics \cite{PES}. However it have been argued that, in effective theories of the SM,  modifications of trilinear gauge boson couplings combined with higher dimensional operators describing fermion gauge boson couplings, are equivalent to oblique contributions from new physics \cite{GRO}.

Our aim is to determine lower bounds for extra neutral gauge boson by a method that takes into account the interaction dynamics involved.  
In order to access the dynamics due to the existence of $Z^\prime$ boson, predicted by two versions of 3-3-1 model, and to obtain new bounds, we have used a specific reaction. We study fermion pair production by considering $Z^\prime$ contribution besides SM $\gamma$ and  $Z$ exchanges in $e^+ e^- $ collisions at ILC \cite{TES} and CLIC \cite{CLI} energies. We explore the range $ M_Z\neq \sqrt s <M_{Z^\prime}$, for which the SM parameters are already precisely known, and we consider the basic process: $e^+ + e^- \to f + \bar{f}$ where $f$ are muons, and quarks $c$ and $b$.

To obtain the lower bounds for $Z^\prime$ mass at $95\%$ C.L., we  perform a $\chi ^{2}$ fit of the differences between the SM and 3-3-1 models calculations for the angular distribution  of one final fermion relative to the initial fermion direction. In order to have realistic results, we consider experimental cuts concerning the detector acceptance.

In the next section we summarize the relevant aspects of two versions of 3-3-1 model. In section III we present our results for energy dependence lower bounds on $Z^\prime$ mass and  $e^+ + e^- \rightarrow e^+ + e^- + e^+ + e^-$ cross section.   Finally, section IV presents  our conclusions.  

\section{Models}

The 3-3-1 models are gauge theories with a largest group of symmetry than SM. They are based on the semi-simple gauge group $SU(3)_C \otimes SU(3)_L \otimes U (1)_X$  and,  as a consequence, they  contain new fermions, scalars and gauge bosons. These particles are until now not experimentally observed, but its existence can lead to interesting signatures. An important motivation to study 3-3-1 models is that the predicted new particles  are expected to occur at energies near the  breaking scale of the SM.

This model offers an explanation of flavor by anomaly cancellation. The  model requires that the number of fermion families be a multiple of the quark color number.  Knowing that
QCD asymptotic freedom condition, is valid only if the number of families of quarks is to be less than five, one concludes that there are three generations.  This offers a possible issue for the flavor problem. 

The electric charge operator is defined in the 3-3-1 model as  
\beq 
Q = T_3 + \beta T_8 + X I
\label {beta} \eq
\nl where $T_3$ and $ T_8$ are two of eight generators satisfying the $SU(3)$ algebra
\beq \left[ T_i\, , T_j\, \right] = i f_{i,j,k} T_k \quad i,j,k =1 .. 8,\eq
\nl  $I$ is the unit matrix and $X$ denotes the $U(1)$ charge before the symmetry breaking.

Electric charge operator  determines how the fields are arranged in each representation and depends on $\beta$.  
Among the choices, $\beta = -\sqrt 3$ corresponds to the minimal version of the model, largely explored in phenomenological applications. The choice $\beta = - 1/\sqrt 3$, which avoids exotic charged fields, leads to a model with right-handed neutrinos. In the following subsections we present the main characteristics of these models.
\subsection{Model I}

In its minimal version, with $\beta=-\sqrt 3$ \cite{PIV,FRA}, the  model has five additional gauge bosons beyond the SM ones. They are: a neutral {$Z'$} and four heavy charged bileptons, ${Y^{\pm\pm},V^\pm} $ with lepton number {$  L = \mp 2$}.  
We display below the lepton content of each generation ($a = 1\dots 3$): 
\begin{eqnarray}
\psi_{a L} = \left( \nu_{a} \ \ell_a \  \ell^{c}_a \right)_{L}^T\ \sim\left({\bf 1}, {\bf 3}, 0 \right), 
 \end{eqnarray} 
\begin{eqnarray} \ell_{a R},\sim \left({\bf 1}, {\bf 1}, -1 \right), \quad \ell^c_{a R} \ \sim \left({\bf 1}, {\bf 1}, 1 \right), 
 \end{eqnarray} 
\nl where $\ell^c_a$ is the charge conjugate of $\ell_a$ ($e$, $\mu$, $\tau$) field. Here the values in the parentheses denote quantum numbers relative to $SU(3)_C$, $SU(2)_L$ and $U(1)_X$.  

Two quark families ($m=1,2$) and the third one are accommodated in $SU(3)_L$ anti-triplet and triplet representation respectively, 
\begin{eqnarray}
Q_{m L} = \left( d_m \  u_m \ j_m
\right)_{L}^T \ \sim \left({\bf 3}, {\bf 3^*}, -1/3 \right),  
\quad Q_{3 L} = \left( u_3 \ d_3 \  J
\right)_{L}^T \ \sim \left({\bf 3}, {\bf 3}, 2/3 \right) 
\end{eqnarray}
\begin{eqnarray}
u_{\alpha R}\ \sim \left({\bf 3}, {\bf 1}, 2/3 \right)&,& \  d_{\alpha R} \ \sim \left({\bf 3}, {\bf 1}, -1/3 \right),\nonumber \\
 J_{R}\ \sim \left({\bf 3}, {\bf 1}, 5/3 \right)&,& \  j_{m R} \ \sim \left({\bf 3}, {\bf 1}, -4/3 \right),
\end{eqnarray} 
\nl $j_1$, $j_2$ and $J$ are exotic quarks with respectively $-4/3$, $-4/3$ and $5/3$ units of positron charge and $\alpha = 1,2,3 $.

The minimum Higgs structure  necessary for symmetry breaking and
that gives quarks and leptons acceptable masses are:
\beq
\eta &=&\left( \eta^{0} \ \eta_{1}^{-} \ \eta_{2}^{+}
  \right)^T \quad \sim \left({\bf 1}, {\bf 3}, 0\right),  
\nn \\
\rho &=&\left(\rho^{+} \ \rho^{0} \ \rho^{++}\right)^T 
  \quad \sim \left({\bf 1}, {\bf 3}, 0\right), \nn
\\ 
\chi &=&\left(\chi^{-} \ \chi^{--} \ \chi^{0}
\right)^T  \quad \sim \left({\bf 1}, {\bf 3}, -1\right).
\nn \\
S &=&\left(\begin{array}{llll}
\sigma^{0}_{1} & \textit{h}^{+}_{2} & \textit{h}^{-}_{1} \\ \textit{h}^{+}_{2} & \textit{H}^{++}_{1} & \sigma^{0}_{2}  \\ \textit{h}^{-}_{1} & \sigma^{0}_{2} & \textit{H}^{--}_{2}  \end{array}\right) \quad \sim \left({\bf 1}, {\bf 6^*}, 0\right), 
\eq
\nl where in parenthesis are the dimensions of $SU(3)_C$ and $SU(3)_L$ representations and the corresponding $U(1)_X$ charges.

The neutral scalars develop non zero vacuum expectation values ($<\chi>$, $<\rho>$, $<\eta>$, and $<S>$) and the breaking of 3-3-1 group to the SM are produced by the following hierarchical pattern
 $${SU_L(3)\otimes U_X(1)}\stackrel{<\chi>}{\longrightarrow}{SU_L(2)\otimes
U_Y(1)}\stackrel{<\rho,\eta, S>}{\longrightarrow}{ U_{e.m}(1).}$$
The consistency of the model with SM phenomenology is imposed by fixing a large scale for the VEV of the neutral $\chi$ field ($v_\chi \gg v_\rho, v_\eta, v_\sigma$), with $v_\rho^2 + v_\eta^2 = v_W^2= \left( 246 \right)^2$ GeV$^2$.

After the breaking, it results a set of gauge bosons: the
standard model  $A, Z$ and $\ W^\pm$ and the new gauge bosons 
$Z^\prime $, $\ V^\pm$ and $Y^{\pm \pm}$.  They
are expressed as a linear combination of $W^a$, $a=1,..8$ and $B$ fields as
\beq
&&W^{\pm}_{\mu}\equiv \frac{W^{1}_{\mu} \mp iW^{2}_{\mu}}{\sqrt{2}},\ 
{    V^{\pm}_{\mu}}\equiv \frac{W^{4}_{\mu} \pm iW^{5}_{\mu}}{\sqrt{2}},\ 
{    Y^{\pm\pm}_{\mu}}\equiv \frac{W^{6}_{\mu} \pm iW^{7}_{\mu}}{\sqrt{2}},
\eq
\beq
A_{\mu}&=& h(t)^{-1/2} \left[ \left(W^{3}_{\mu} - \sqrt{3}\, W^{8}_{\mu}\right)t + B_{\mu} \right],\cr
\cr
Z_{\mu}&\simeq& - h(t)^{-1/2} \left[ f(t)^{1/2} W^{3}_{\mu} +
 f(t)^{-1/2}\left(\sqrt{3}\,t{^2}\, W^{8}_{\mu} - t B_{\mu}\right)
 \right],\cr \cr
{Z^\prime_{\mu}}&\simeq & f(t)^{-1/2}\left[W^{8}_{\mu} + \sqrt{3}\, t\, B_{\mu}\right],
\eq
\nl with\, $t = g^\prime / g$, we write  $h(t)=1+4 t^2$\,  and\,  $f(t)=1+3 t^2$.
\bigskip

The charged gauge boson masses as a function of VEV's are:
\begin{eqnarray}
M^2_{W}=\frac{1}{4} g^{2}\left({v}_{\eta}^{2} + {v}_{\rho}^{2}\right),
M^2_{V}=\frac{1}{4} g^{2}\left({v}_{\rho}^{2} + {v}_{\chi}^{2}\right),
M^2_{Y}=\frac{1}{4} g^{2}\left({v}_{\eta}^{2} + {v}_{\chi}^{2}\right).
\end{eqnarray}

and the neutral gauge boson masses 

\begin{eqnarray}
M^2_{\gamma}=~0,~M^2_{Z} \simeq ~\frac{g^{2}}{4}\, \frac{h(t)}{f(t)} \left(v^{2}_{\eta} +
v^{2}_{\rho}\right),~M^{2}_{Z^{\prime}} \simeq~ \frac{g^{2}}{3}\, f(t)\,  v^{2}_{\chi}
\end{eqnarray}

The VEV's induce $Z$-$Z^{\prime}$ mixing. As a consequence the physical states $Z_1$ and $Z_2$ are related to $Z$ and $Z^{\prime}$ states by a mixing angle,
\beq
\tan^2 \theta = \frac{M^2_Z - M^2_{Z_1}} {M^2_{Z_2} - M^2_Z},
\eq
\nl where $Z_1$ corresponds to SM neutral gauge boson and $Z_2$ to the extra neutral one. 
For a small mixing, $\theta \ll 1$, $Z_2$ corresponds to $Z^\prime$.

The relation between $Z^\prime$ and $Y$ masses \cite{DION,NGL}, in the minimal model is:
\beq
&&{M_{Y}\over M_{{Z^{\prime}}}} \simeq {M_{V}\over M_{{Z^{\prime}}}} \simeq {\sqrt{3-12\sin^2\theta_W}\over {2\cos\theta_W}}. 
\eq
\nl From this relation and using $\sin^2\theta_W =0.23$ \cite{PDG} the ratio becomes $\simeq 0.3 $ so that $Z^{\prime}$ can decays in a bilepton pair.
   
The neutral interactions involving leptons follow from
\beq
{\cal L}^{NC} = - \bar \ell \, \gamma^\mu \, \ell\  A_\mu  -  \frac{g}{4}\frac{M_Z}{M_W}[\bar
\ell\, \gamma^\mu\ (v_\ell+a_\ell\gamma^5)\ \ell\, Z_\mu+ \bar
\ell\, \gamma^\mu\ (v^\prime_\ell+a^\prime_\ell\gamma^5)\ \ell\, { Z_\mu^\prime}],
\eq
\nl with
$ v_\ell =   -1/h(t),\ a_\ell = 1, \quad
v^\prime_\ell =  -\sqrt{3/h(t)},\quad 
\  a^\prime_\ell= v^\prime_\ell/3, $

\nl Note that for $t^2=11/6$, $v_\ell$ and $a_\ell$ have the same values as the SM couplings. One of the main features of the model comes from the relation between the  $SU_L(3)$  
  and $U_X(1)$ couplings, 
($g^\prime/g =t$)   expressed as:
\begin{equation}
\frac {g^{\prime\, 2}}{g^2} =\frac{\sin^2 \theta_W}{1\, -\, 4 \sin^2 \theta_W}.
\end{equation}
that fixes $\sin^2 \theta_W < 1/4$, which is a peculiar characteristic of this model.
\subsection{Model II}

Another model considered in the present work, corresponds  to the choice $\beta = -1/\sqrt 3$   in the Eq.(\ref {beta}) and adds an anti-neutrino to each $SU(2)_L$ SM  lepton doublet to form a $SU(3)_L$ triplet \cite{RHN}
\beq
\Psi_{a L} = \left( 
               \nu_a,\  e_a,\  \nu^C_a
 \right)_L^T \sim ({\bf 1}, {\bf 3}, -1/3),\ \  e_{a R}\sim ({\bf 1},
{\bf 1}, -	1), 
\eq
 where $a=1,2,3$ is the generation index.

The two first quark generations ($m=1,2$)  belong to  anti-triplets and the third one
to triplet representation 
 \beq Q_{m L} = \left(d_{m},\  u_{m},\   d^{\prime}_{i} \right)_L^T \sim ({\bf 3}, {\bf 3^*}, 0), 
\ \ 
 Q_{3L} = \left( u_{3}, \ d_{3}, \ u^\prime_3 
                 \right)_L^T \sim ({\bf 3}, {\bf 3}, 1/3),
\eq
\beq
 u_{\alpha R} \sim ({\bf 3}, {\bf 1}, 2/3),  \ \  d_{\alpha R}\sim ({\bf 3}, {\bf 1}, -1/3),\nn \\
 u_{3 R}^\prime  \sim ({\bf 3}, {\bf 1}, 2/3), \ \ d^\prime_{m R}\sim ({\bf 3}, {\bf 1}, -1/3).
\eq

For symmetry breaking one introduces three triplets:
\beq
\eta &=&\left( \eta^{0}, \ \eta^{-}, \ \eta^{\prime\  0}
  \right)^T \ \  \sim \left({\bf 1}, {\bf 3}, -1/3\right),  
\nn \\
\rho &=&\left(\rho^{+}, \ \rho^{0}, \ \rho^{\prime \ +}\right)^T 
  \ \  \sim \left({\bf 1}, {\bf 3}, \ 2/3\right), \nn
\\ 
\chi &=&\left(\chi^{0}, \ \chi^{-}, \ \chi^{\prime\  0}
\right)^T  \ \  \sim \left({\bf 1}, {\bf 3}, -1/3\right). 
\eq

The model has five neutral scalars. Similarly to Model I, one can consider that only three of them develop non zero VEV:  $\langle \eta^0 \rangle = v_\eta/\sqrt 2$, \ $\langle \rho^0 \rangle = v_\rho/\sqrt 2$, \ $\langle \chi^{\prime \ 0} \rangle = v_{\chi}^\prime/\sqrt 2$, with $v_\chi \gg v_\rho, v_\eta$, in order to reproduce the SM low energy phenomenology.

 The consequences to consider more than three non zero VEV are: leptonic number violation, Majorana neutrinos mass generation and the existence of a Goldstone boson, called Majoron. 

One of the main features of the model comes from the relation between the  $SU_L(3)$  
coupling, $g$,  and $U_X(1)$ coupling, $g^{\prime}$ ($g^\prime/g =t$)  expressed as:

\beq
\frac {g^{\prime \, 2}}{g^2} =\frac{2\sin^2\theta_W}{1 - 4/3 \sin^2\theta_W}.
\eq

The gauge bosons are $W^{a}_{\mu}$ ($a=1... 8$) in a octet representation of  ${SU(3)_{L}}$ and a singlet $B_{\mu}$ of $U(1)_{X}$.   
The charged and neutral  gauge bosons are defined from the combinations:
\beq
&&W^{\pm}_{\mu}\equiv \frac{W^{1}_{\mu} \mp iW^{2}_{\mu}}{\sqrt{2}},\ { V^{\pm}_{\mu}}\equiv \frac{W^{6}_{\mu} \pm iW^{7}_{\mu}}{\sqrt{2}},\  X^0_{\mu} = \frac{W^{4}_{\mu} - iW^{5}_{\mu}}{\sqrt{2}},
\eq
\beq
A_{\mu}&=& h^{\prime}(t)^{(-1/2)}\left[\frac{t}{\sqrt{2}}\, \left( \sqrt{3} \, W^{3}_{\mu} - W^{8}_{\mu}\right)  + 3\,B_{\mu} \right],\cr
Z_{\mu}&\simeq &  {h^\prime(t)^{-(1/2)}\, f^\prime(t)^{-(1/2)}} \left[ \frac {f^\prime(t)} {\sqrt{2}} \, W^{3}_{\mu} -\, \frac{\sqrt{3}} {2}\, t^2\, W^{8}_{\mu} + 
3 \sqrt{3}\, t B_{\mu}  \right],\cr
{Z^\prime_{\mu}}&\simeq & f^\prime(t)^{-(1/2)} \left[ 3\sqrt {2}\, W^{8}_{\mu} + \, t\,  B_{\mu}\right],
\eq

\nl with\, $t = g^\prime/g$, we write  $h^\prime(t)= 9 +2 t^2 $ \, and \,   $f^\prime (t)= 18 + t^2 $ .
\bigskip

The charged gauge boson $W$, $V$ and the neutral $X$ masses as a function of VEV's are:

\begin{eqnarray}
M^2_{W}=\frac{1}{4} g^{2}\left({v}_{\eta}^{2} + {v}_{\rho}^{2}\right),
M^2_{V}=\frac{1}{4} g^{2}\left({v}_{\rho}^{2} + {v}_{\chi}^{2}\right),
M^2_{X}=\frac{1}{4} g^{2}\left({v}_{\eta}^{2} + {v}_{\chi}^{2}\right).
\end{eqnarray}

and the photon, $Z$  and $Z^{\prime}$ masses are:
 
\begin{eqnarray}
M^2_{\gamma}&=&~0, \cr 
~M^2_{Z} &\simeq& ~\frac{g^{2}}{2} \frac{h^\prime(t) }{f^\prime (t)}\left (v^{2}_{\eta} + v^{2}_{\rho}\right),\cr \nonumber \\
~M^{2}_{Z^{\prime}} &\simeq& \frac{g^2}{54\, f^\prime(t)}\Bigl[h^\prime(t)^2 v^{2}_{\rho}+ \left( t^2 - 9 \right)^2 v^{2}_{\eta}+ f^\prime(t)^2 \, v^{2}_{\chi}\Bigr].
\end{eqnarray}

The relation between $Z^\prime$, $V$ and $X$ masses \cite{DION,NGL}, in right-handed neutrino model model is:
\beq
&&{M_{V}\over M_{{Z^{\prime}}}}\simeq {M_{X}\over M_{{Z^{\prime}}}}\simeq{{\sqrt{3-4\sin^2\theta_W}}\over {2\cos\theta_W}}, 
\eq
\nl which will be used in the present work. From this relation we obtain $\simeq 0.82$ and we can see that $Z^\prime$ is forbidden to  decay in a bilepton pair ($V \bar V$ or $X \bar X$).
\sk

The neutral interactions involving leptons follow from the Lagrangian
\beq
{\cal L}^{NC} = - \bar \ell \, \gamma^\mu \, \ell\  A_\mu  -  \frac{g}{c_W}[\bar
\ell\, \gamma^\mu\ (v_\ell+a_\ell\gamma^5)\ \ell\, Z_\mu+ \bar
\ell\, \gamma^\mu\ (v^\prime_\ell+a^\prime_\ell\gamma^5)\ \ell\, { Z_\mu^\prime}],
\eq
\nl with
$ v_\ell = \displaystyle{-\frac {1} {4}} +\sin^2 \theta_W $, \,   $a_\ell = \displaystyle{\frac {1} {4}}$,  \,
$v^\prime_\ell = \displaystyle{\frac{v_\ell}{\sqrt { 3-4\sin ^2 \theta_W}}}$, \, 
$a^\prime_\ell= \displaystyle{\frac{a_\ell}{\sqrt { 3-4\sin ^2 \theta_W}}}$,
\nl where $\ell = e,\,  \mu, \, \tau $.

\section{Results}

Several approaches have been used to obtain bounds on $Z^\prime$ mass and mixing angle, for some versions of 3-3-1 model. The most explored, among them, calculates the oblique electroweak correction parameters ($S$, $T$ and $U$) due to exotic particles contribution \cite{LIU,SAS}. Other direct approaches  uses the  experimental results at $Z$-pole generalized for arbitrary
$\beta$ values \cite{OCH,FRE,GUT} 
or the contributions of $Z^\prime$ to neutral mesons ($K$, $D$ and $B$) mass difference due to the flavor changing neutral current (FCNC) \cite{VAN,DUM,TAE}. The bounds on $Z^\prime$ mass were also obtained by considering the energy region were perturbative treatment is still valid \cite{ALE,PHF}.  
An indirect method to establish limits on $Z^\prime$ mass follows from the relation between $Z^\prime$ and exotic boson masses, and it consists in the analysis of exotic boson contribution to the muon decay parameters \cite{NGL,BEL}. Besides, another model, with lepton families in different representations, obtained bounds for $Z^\prime$ mass from $\mu \rightarrow 3 \ e $ \cite{SHE}.

In this paper we obtain new bounds on the $Z^\prime$ mass from the collision $e^+ + e^- \longrightarrow  f + \bar f$, with $f = \mu$, or quark $c$ or $b$. This process has the advantage that, at tree level, it introduces  few unknown parameters since there is no contribution from bileptons or other exotic model particles. Apart from  SM contributions ($\gamma$ and $Z$), the only new contribution comes from s-channel $Z^\prime $ exchange.
 Another inherent advantage  by considering scattering process is that it gives energy dependent bounds. On the other hand, the consistence of our analysis is based on the choice of two different quark content but with the same group representation for both models: two anti-triplets for the first and second families and one triplet for the third one.
 
Our strategy is to perform a $\chi^2$ analysis of the difference between the angular distribution of the final fermions relative to the initial beam, predicted by the SM, with those from the two versions of the 3-3-1 model.
The procedure is as follows: supposing that the experimental data for the fermion pair production to be described by the SM, we define a one-parameter $\chi^2$ estimator
\smallskip
\begin{equation}
{ \chi^2 = \sum_{i=1}^{n_b} {\biggl( {N_i^{SM}- N_i^{331} \over
\Delta N_i^{SM}}\biggr)^2}},
\end{equation}
where $N_i^{SM}$ is the number of SM events collected in the
$i^{th}$ bin, $N_i^{331}$ is the number of events in the $i^{th}$ bin as
predicted by the considered model, and $\Delta N_i^{SM} =
\sqrt{(\sqrt {N_i^{SM}})^2 + (N_i^{SM}\epsilon)^2}$ the corresponding
total error, which combines in quadrature the Poisson-distributed statistical
error with the systematic error.
 We took $\epsilon = 5\%$ as the systematic error
in our calculation. We considered the muon, charm and bottom detection efficiency  as $95\%$, $60\%$ and $35\%$ respectively.

To obtain $Z^\prime$ total width, we take $M\simeq 600$ GeV for the exotic quarks and, for bileptons masses, we keep  the constraints with $M_{Z^\prime}$
given by Eqs. (13) and (25) for Models I and II respectively. 
For Model II it results in a very narrow resonance,  while for Model I it appears broader, as can be seen in Figure 1. This scenario does not change, when we consider 
exotic quarks masses equal to $1$ TeV, as can be seen in Figure 2. This happens because, in Model I, it is allowed for $Z^\prime$ to decay in bileptons. We can also observe that, only for Model I, there is a rapidly grow of $\Gamma_{Z^\prime}$ related to a big branching ratio for exotic quark pair production.

Using the three above refereed final channels,
we estimated bounds for $M_{Z^\prime}$ with $95\%$ C.L. for energy range from  $\sqrt s$ from $0.5$ TeV to $3$ TeV corresponding to the proposed ILC and CLIC experiments. Our results for bounds from  $\bar \mu \mu$,  $\bar c c$ and  $\bar b b$ channels, are displayed in Figures 3, 4 and 5 respectively. 

For Model I, Eq. (15)  leads to $\sin^2 \theta_W < 1/4$ (Landau pole). The energy scale corresponding to the Landau pole gives an upper bound for $M_{Z^\prime} \simeq 5$ TeV \cite{ALE}. This bound is represented by horizontal lines in Figures 3, 4 and 5. Landau pole analysis do not impose any upper bound for Model II.

\begin{figure}[!htb] 
\includegraphics[height=.5\textheight]{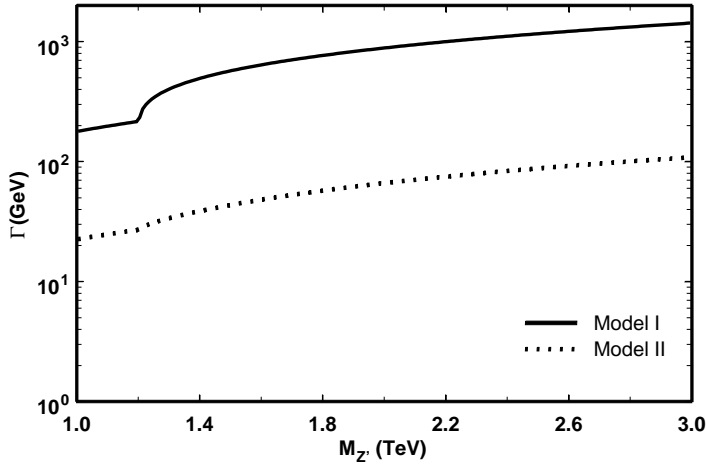} 
\vskip -1.2cm
\caption{The $Z^\prime$ total width for Model I and Model II (considering exotic quark masses of the order of 600 GeV).}
\end{figure}

\begin{figure}[!htb] 
\includegraphics[height=.5\textheight]{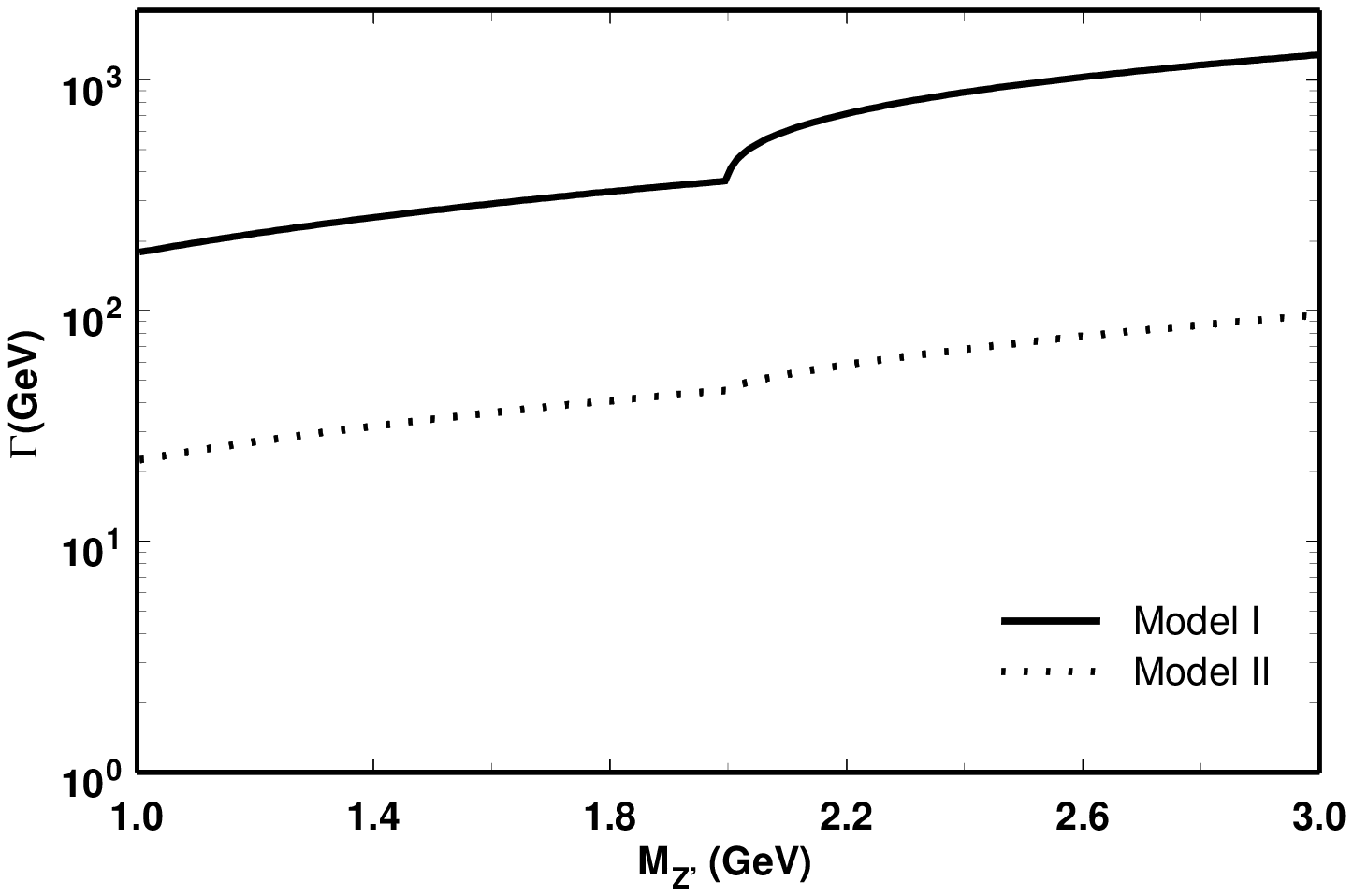}
\vskip -1.2cm
\caption{The $Z^\prime$ total width for Model I and Model II (considering exotic quark masses of the order of 1 TeV).}
\end{figure}

\begin{figure}[!htb] 
\includegraphics[height=.5\textheight]{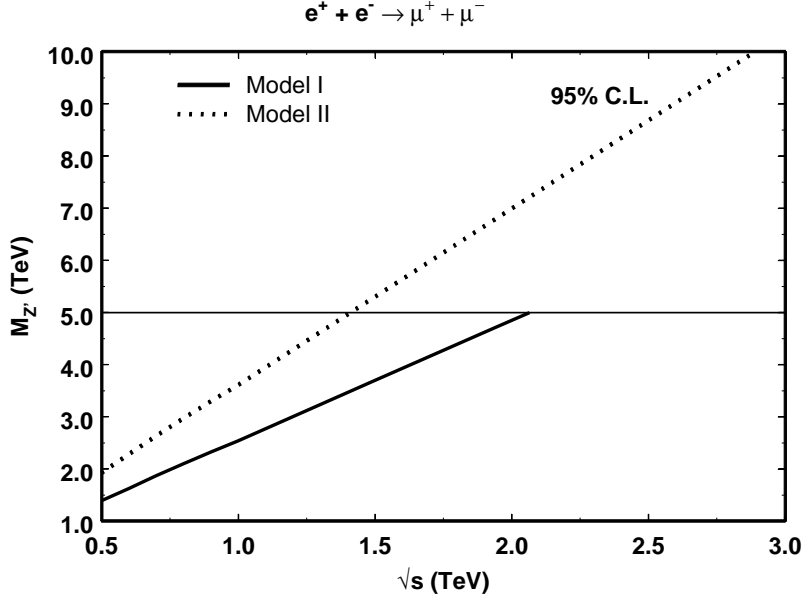}
\vskip -1.2cm
\caption{The lower bounds at $95 \% $ C.L. extracted from the angular distribution of one muon relative to the beam direction {\it versus} $\sqrt s$, for $e^+ + e^- \rightarrow \mu^+ + \mu^-$ for Models I and II.}
\end{figure}

\begin{figure}[!htb] 
\includegraphics[height=.5\textheight]{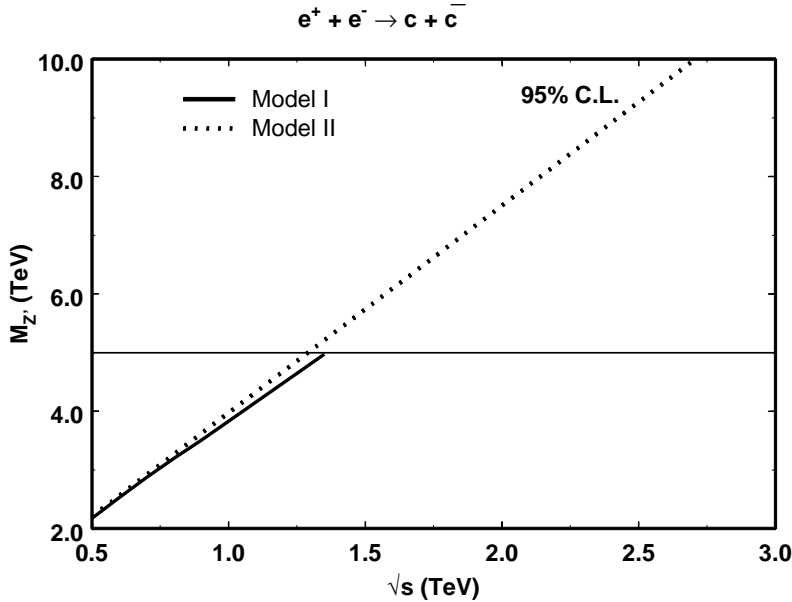}
\vskip -1.2cm
\caption{The lower bounds at $95 \% $ C.L. extracted from the angular distribution of one charm quark  relative to the beam direction {\it versus} $\sqrt s$, for $e^+ + e^- \rightarrow c + \bar c$ for Models I and II.}
\end{figure}

\begin{figure}[!htb] 
\includegraphics[height=.5\textheight]{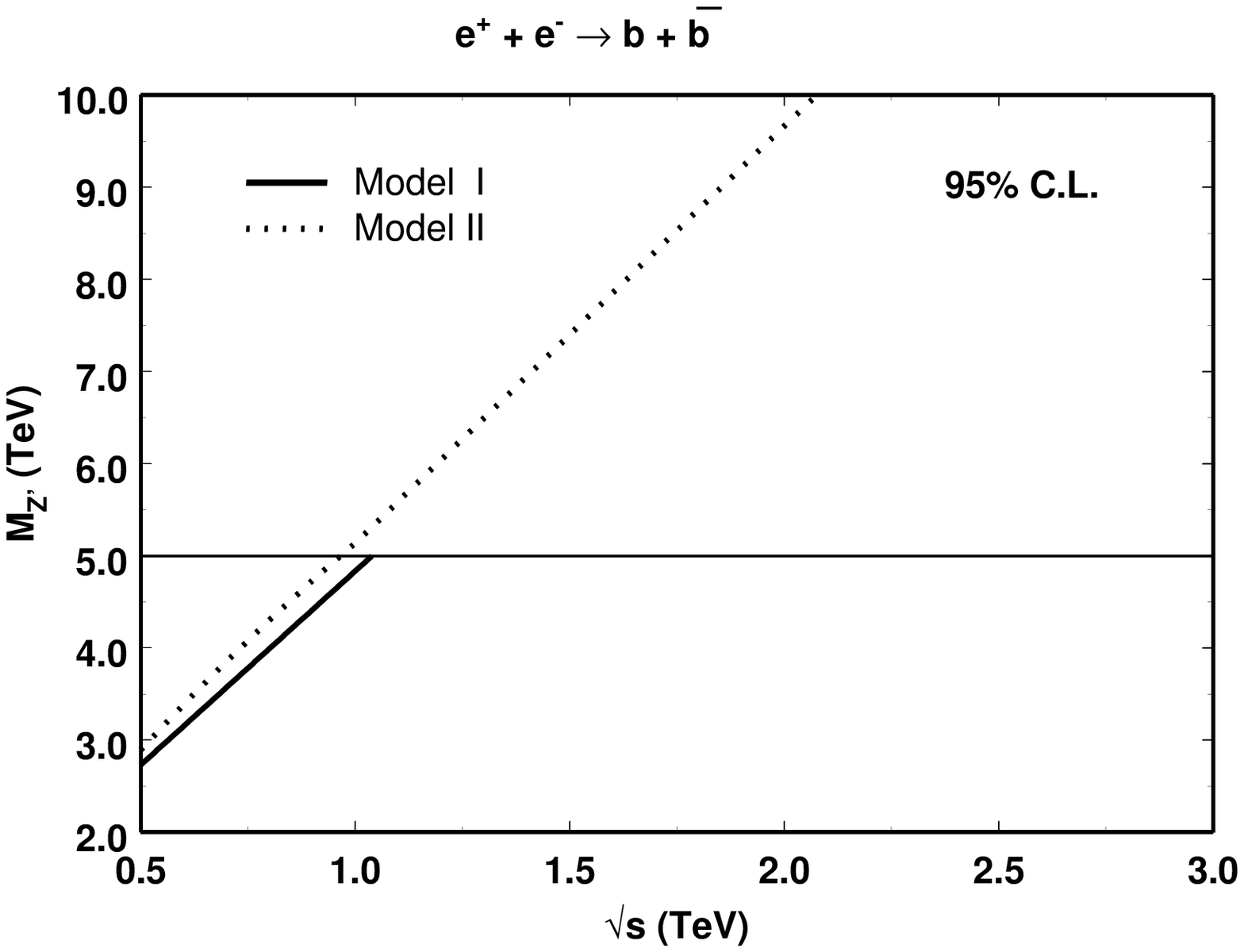}
\vskip -1.2cm
\caption{The lower bounds at $95 \% $ C.L. extracted from the angular distribution of botton quark  relative to the beam direction {\it versus} $\sqrt s$, for $e^+ + e^- \rightarrow b +\bar b$ from Models I and II.}
\end{figure}

We observe that Model I gives lower bounds values than Model II for the three considered channels. Besides, lepton channel is less restrictive than the hadron channels. Having in mind the bound given by Landau pole analysis,
we display in Table I our results for $M_{Z^\prime}$ for $\sqrt s= 0.5$ TeV and
$\sqrt s=3$ TeV for both models. For Model I the maximum value of $M_{Z^\prime}$ is around  $5$
TeV, but for Model II  $M_{Z^\prime}$ can be greater than $10$ TeV. This result is usefull to disentangle the predictions from these 3-3-1 versions. We resume in Table II, the bounds obtained by other approaches. We conclude that there is a complete agreement between our energy dependent results with all bounds present in the Table II. 

To be more specific let us compare our results, with those obtained from an exhaustive analysis done in Ref. \cite{OCH}. These authors do a $\chi^2$ fit with $95 \%$ C.L.  between LEP parameters at the $Z$-pole  and the predictions from three different quark representations of 3-3-1 models ($A$, $B$ and $C$). They consider three degrees of freedom: the $Z^\prime$ mass, the mixing angle $\theta$ between $Z$ and $Z^\prime$ mass states and the $\beta$ parameter. 

In our case, we work in the representation A,  with $\beta=-\sqrt 3$ (Model I) and  $\beta=-1/\sqrt 3$ (Model II) and we do not consider any mixing between $Z$ and $Z^\prime$ mass states. Below we present the reasons why our results, that covers a wide energy range,   are in complete agreement with their constraints:
\begin{itemize}
\item  Model I and II do not allow $M_{Z^\prime} < 1$ TeV,
\item  Model I is excluded for $M_{Z^\prime} < 1.5$ TeV,
\item  both Models allow masses larger than $2$ TeV.
\end{itemize} 

Ref. \cite{OCH} displays the allowed region in $\sin \theta \times \beta $ plane for some $Z^\prime$ masses. The allowed region always includes our zero mixing choice. They also present an allowed region in $M_{Z^\prime}\times \beta $ plane for  $10 ^{-3}> \sin \theta >10 ^{-4}$. In spite of their small mixing, once again, our results are compatible with their bounds.  

\begin{table}[h]\label{xuxa}
\begin{footnotesize}
\begin{center}
\begin{tabular}{||c|c|c|c|c||}
\hline \hline

\multicolumn{3}{|c|}{$\sqrt s= 0.5$ TeV} &
\multicolumn{2}{|c|}{$\sqrt s=  3$ TeV} \\ \hline
\hline
&    &     &    &      \\
&  Model I & Model II   & Model I & Model II \\
&    &    &    &       \\ \hline
\hline
&    &     &    &      \\
$\mu^+ + \mu^-$ &  $1.5$ TeV & $2$ TeV & $-$ & $10$ TeV \\
&     &    &    &    \\ \hline
\hline
&    &     &    &      \\
$c + \bar c$ &  $2$ TeV & $2$ TeV & $-$ & $12$ TeV\\
&     &    &    &    \\ \hline
\hline
&    &     &    &      \\
$b + \bar b$ &  $2.5$ TeV & $2.5$ TeV & $-$ & $15$ TeV \\
&     &    &    &    \\ \hline
\hline
\end{tabular}
\end{center}
\end{footnotesize}
\caption{Lower bounds for $M_{Z^\prime}$ for lepton and hadron channels at $\sqrt s= 0.5 $ 
TeV and $\sqrt s= 3$ TeV from Models I and II.}

\end{table}
\par

\begin{table}[t!]\label{morango}
\begin{center}
\begin{tabular}{||c|c|c||}
\hline \hline
     &   &    \\
 Approaches &  $M_{Z^\prime}$ (Model I) & $M_{Z^\prime}$ (Model II) \\ \hline \hline
     &   &    \\ 
$S$, $T$, $U$  & $1.4$- $2.2$ TeV$^{a,b}$ & $> 1.2$ TeV \\
    &     &   \\ \hline
\hline
    &      &  \\
$Z$-pole  &  $> 1.8$ TeV$^{c,d}$ & $> 1.8$ TeV$^{c,d} $, $>2.1$ TeV$^{e} $\\
    &     &    \\ \hline
\hline
    &     & \\
Landau pole    & $0.8$-$5.2$ TeV$^f$, $> 1.4 $ TeV$^g$  &  -\\
    &     &   \\ \hline
\hline
    &     & \\
$\sin^2\theta_W$    & $1.3$-$3.1$ TeV$^h$ & - \\
    &     &   \\ \hline
\hline
    &     & \\
Muon decay    & $1.3$-$3.1$ TeV$^{h,i}$ & - \\
    &     &  \\ \hline
\hline
    &     & \\
FCNC    & $1.8$-$3.1$ TeV$^j$ , $< 2$ TeV$^k$ & $> 1 $ TeV$^l$ \\
    &     &  \\ \hline
\hline
\end{tabular}
\end{center}
\caption{ Predicted bounds for $M_{Z^\prime}$ from different approaches: $a$ \cite{LIU},\ $b$ \cite{SAS},\  $c$ \cite{OCH},\ $d$ \cite{FRE},$e$ \cite{GUT}, \ $f$ \cite{ALE}, \ $g$ \cite{PHF}, \ $h$ \cite{NGL}, \ $i$ \cite{BEL}, \ $j$ \cite{TAE} , \ $k$ \cite{DUM}, \ $l$ \cite{VAN}.}
\end{table}

\par
One interesting byproduct of our analysis concerns the possibility for disentangle the proposed models. 
Let us apply the obtained bounds to go further in this analysis.  It is clear that for $\sqrt s < 4$ TeV there is no available phase space to produce a pair of on-shell $Z^\prime$ with $M_{Z^\prime} = 2$ TeV. However, it is important to notice that the way we estimate $Z^\prime$ mass relies on several arbitrariness: statistical errors, systematic errors, detection efficiency, bin number selection, angular and energy cuts and so on. This way, in the following, we take a less restrictive value for $Z^{\prime}$ mass. 
For example, considering $M_{Z^\prime}$ just below our bounds, say $1.5$ TeV, the resulting cross sections for $e^+ + e^- \longrightarrow Z^\prime + Z^\prime$ are displayed in Figure 4,  that also includes the dominant SM contribution for two $Z$ production. From this graph, it is clear that only CLIC machine can produce a pair of such heavy neutral boson.  
For an annual integrated luminosity ${\cal L}_{int}=  500$ fb$^{-1}$ the number of events with two final $Z^\prime$ is  ${\cal O} (10^{2})/yr$  for 3-3-1 models and ${\cal O} (10^{4})/yr $ for two final $Z$ in SM.

\begin{figure}[!htb] 
\includegraphics[height=.5\textheight]{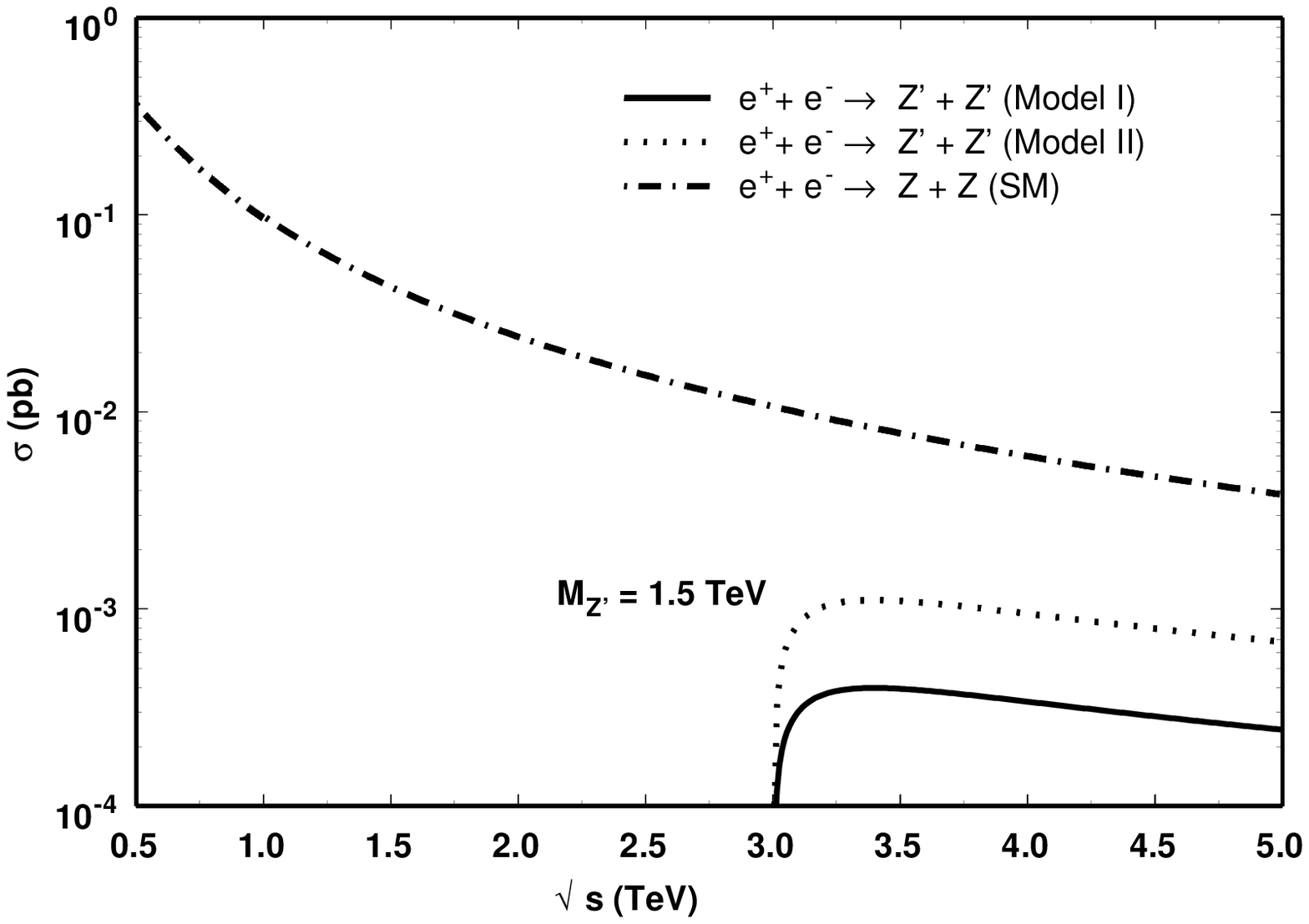}
\vskip -1.2cm
\caption{Total cross section for $e^+ + e^-$ collisions producing two neutral gauge boson
against $\sqrt s$ for Models I, II and SM}
\end{figure}

In order to look at a possible signature of $Z^\prime$, we extend our analysis to $e^+ + e^- \rightarrow e^+ + e^- + e^+ + e^-$ reaction within the same previous hypothesis. In the Model I, appart from $Z^\prime$ contribution, the new physics comes from double charged bileptons $Y^{\pm\pm}$. In this model, a copious production of two pairs of same-sign leptons is expected to occur due to the large  bilepton-leptons coupling. In our recent work \cite{PLB}, we obtained:
$10^2- 10^4$ events for $\sqrt s= 1$ TeV ($276$ GeV $< M_Y< 500$ GeV and $1$ TeV $ < M_{Z^\prime}<1.8$ TeV) and $10^3$ events ($500$ GeV $< M_Y< 800$ GeV and $2$ TeV $ < M_{Z^\prime}<3$ TeV) for $\sqrt s= 3$ TeV. 
On the other hand, for Model II, the new physics for four fermions production comes only associated with $Z^\prime$. 
The number of tree diagrams increases from 36 corresponding to SM contributions to  81   diagrams for Model II. To obtain
cross sections and distributions we use  CompHep package \cite{HEP} due the complexity of the calculation.

For the detector acceptance, we adopted an angular cut of $\vert \cos \theta \vert \le 0.995 $ for the direction of final leptons relative to the beam, and energy cut of $5$ GeV for  final leptons \cite{MIK,AZU,L3C}. To show the signature for $ Z^{\prime}$ existence, we have also selected an invariant mass cut for final pair ($e^+ e^-$),  $\vert M_{ee}-M_{Z^{\prime }}\vert \simeq \Gamma_{Z^{\prime}}$, where the $\Gamma_{Z^{\prime}}$ widths are presented in Figures 1 and 2.

For CLIC energy bigger than $3$ TeV, it would be possible to produce four leptons  from a pair of $Z^{\prime}$ with mass equal $1.5$ TeV each. In this conjecture we obtain a total cross section $\sigma \simeq 10^{-6}$ pb, corresponding to a maximum 0.5 events/yr, for  ${\cal L}_{int}=  500$ fb$^{-1}$. Such small cross section shows that this process is not the best ground to study the contribution from $ Z^{\prime}$ in 3-3-1 models. On the other hand, four lepton production in linear collider gives a good signature for the bilepton existence as we have shown in a previous work for Model I \cite{PLB}.  

\section{Conclusions}
Working with two versions of 3-3-1 model, the minimal version and another that allows right-handed neutrinos, we explore energy ranges accessible in next linear colliders (ILC and CLIC) , $\sqrt s = 0.5$ TeV to $5$ TeV. To obtain the lower bounds for $M_{Z^\prime}$,  we perform  
a $\chi ^{2}$ fit of the difference between SM final lepton angular distribution and that predicted by the 3-3-1 models at a given energy in the process $e^+ + e^- \longrightarrow f^+ +  f^-$, with $f=\mu, c $ and $b$. The  energy dependence of the bounds was determined within $95\%$ C.L. Our main result indicates that the $Z^\prime$ mass is larger than  $2$ TeV, in complete agreement with previous predictions shown in Table II. 

In addition, we show that, as the angular distributions depends on $Z^\prime$ couplings,  the obtained bounds allow one to disentangle the studied models. We observe that the final channel with muons is less restricted than the hadron channel ones. For a unique representation for Models I and II, small $Z^\prime$ masses are excluded, and the bounds for Model I are lower than the ones from Model II.

It is interesting to note that angular distributions are tools to reveal contributions beyond SM only for large values of $M_{Z^\prime}$. In this work, the lower bounds obtained  reflects this fact.  On the other hand, in a recent paper \cite{AMP}, it was proved that 3-3-1 models can be realized also at low energy scales, showing up the effects of the new physics near the electroweak scale ($M_{Z^\prime} \simeq 300$ GeV). However, angular distributions are not sensitive to new particles with small mass. In principle, the contributions to oblique corrections ($S$, $T$, $U$) and the dependence of some cross sections with the VEV's, can help to explore the proposed low energy scale for new physics \cite{MON}.   

Appart from our main proposal to obtain energy dependent bounds on $Z^\prime$ mass,  we have also  performed a complete tree level calculation of four leptons production mediated by   $Z^\prime$ in Model II, where there is no bilepton contribution. To do that, we applied angular and energy cuts for the detector acceptance and we introduced cuts in invariant mass and transverse momenta of final leptons to distinguish the signal from SM background. For  $M_{Z^{\prime}} = 1.5$ TeV and $\sqrt{s}> 3$ TeV, we obtained less than  one event by year.  

We conclude that this process, when calculated in the model with right-handed neutrinos,   does not give a clear signature for $Z^\prime$ production at CLIC. We intend to apply the present procedure for other possible 3-3-1 model versions like \cite{SHE}. 
   
\vskip 1cm
We acknowledge the financial support from CAPES (E.~R.~B.~) and FAPERJ (Y.~A.~C.~).

\ed